\documentclass[prl,aps]{revtex4}

\usepackage{epsfig}
\usepackage{amsfonts}
\usepackage{rotating} 
\usepackage{sparticles}
\usepackage{dcolumn}
\usepackage{bm}
\usepackage{hyperref}
\hypersetup{
    bookmarks=true,         
    unicode=false,          
    pdftoolbar=true,        
    pdfmenubar=true,        
    pdffitwindow=false,     
    pdfstartview={FitH},    
    pdftitle={My title},    
    pdfauthor={Author},     
    pdfsubject={Subject},   
    pdfcreator={Creator},   
    pdfproducer={Producer}, 
    pdfkeywords={keyword1} {key2} {key3}, 
    pdfnewwindow=true,      
    colorlinks=true,       
    linkcolor=red,          
    citecolor=cyan,        
    filecolor=magenta,      
    urlcolor=green,           
    linktocpage=true
}
\usepackage{amsmath,amssymb,amsthm,graphicx,latexsym}
\usepackage{subfigure}
\usepackage{pstricks}
\usepackage{color}

\usepackage{mathrsfs}

\newcommand{\be}{\begin{equation}}
\newcommand{\ee}{\end{equation}}
\newcommand{\bea}{\begin{eqnarray}}
\newcommand{\eea}{\end{eqnarray}}
\newcommand{\bml}{\begin{subequations}}
\newcommand{\eml}{\end{subequations}}
\newcommand{\bfig}{\begin{figure}}
\newcommand{\efig}{\end{figure}}

\newcommand{\slashed}{\hspace{-1.1ex}/}

\begin{document}

\title{Galileogenesis: A new cosmophenomenological zip code for reheating through R-parity violating coupling}

\author{{{Sayantan Choudhury}}$^{1}$\footnote{Electronic address: {sayanphysicsisi@gmail.com}} ${}^{}$
 and {{Arnab Dasgupta}}$^{2}$
\footnote{Electronic address: {arnab@ctp-jamia.res.in
}} ${}^{}$}
\affiliation{$^1$Physics and Applied Mathematics Unit, Indian Statistical Institute, 203 B.T. Road, Kolkata 700 108, INDIA
}
\affiliation{$^2$Centre for Theoretical Physics, Jamia Millia Islamia, Jamia Nagar, New Delhi-110025, INDIA
}

\begin{abstract}
In this paper we introduce an idea of leptogenesis scenario in higher derivative gravity induced DBI Galileon framework {\it aka Galileogenesis} in presence of one-loop 
R-parity violating couplings in the background of a low energy effective supergravity setup
 derived from higher dimensional string theory framework. We have studied extensively
the detailed feature of reheating constraints and the cosmophenomenological consequences of thermal gravitino dark matter 
in light of PLANCK and PDG data. Finally we have also established a direct cosmophenomenological connection among dark matter relic abundance, reheating temperature
and tensor-to-scalar ratio in the context of DBI Galileon inflation.

\end{abstract}


\maketitle


The post big bang universe
passed through various phases in which reheating plays the crucial
role in explaining production of different particle species from
inflaton. Particle cosmologists have a clear picture
of this hot big bang phase because ordinary matter and
radiation were driving it and also the physical processes
that characterize it involve terrestrial physics.
These particles interact with each other and eventually
they come to a state of thermal equilibrium. This process completes when all
the energy of the inflaton transfer to the thermal energy of elementary particles. 
Amongst all particles degrees of freedom the production of thermal gravitinos during 
reheating \cite{Allahverdi:2007zz,Allahverdi:2010xz,Allahverdi:2006we,Cardenas:2007xh,Boyanovsky:1995ema,Kofman:1997yn,McDonald:1999hd} and its decay play a pivotal role in the
context of leptogenesis \cite{Fong:2013wr,Okada:2005kv,Pradler:2006qh,Rangarajan:2006xg} and dark matter detection \cite{Mazumdar:2010sa,Enqvist:2003gh,Jungman:1995df,Pradler:2006hh}. In a most general prescription usually two types
of gravitinos are produced in this epoch - stable and
unstable. Both of them stimulate the light element
abundances during BBN \cite{Pospelov:2010hj,Burles:2000ju,Kohri:2005wn} and directly affects
the expansion rate of the universe. The gravitino energy density is proportional to gravitino abundance which is obtained by
considering gravitino production in the radiation dominated era following reheating \cite{Kallosh:1999jj,Maroto:1999ch,Bolz:2000fu}. 

In this paper we perform our complete phenomenological analysis with a potential driven DBI Galileon 
framework in the background of ${\cal N}=1,{\cal D}=4$ supergravity
 \cite{Nilles:1983ge,Choudhury:2011sq,Choudhury:2011rz,Choudhury:2012ib,Choudhury:2013jya,Choudhury:2013zna,Choudhury:2014sxa,Choudhury:2014uxa,Mazumdar:2011ih} which can be obtained from the
dimensional reduction from higher dimensional string theory setup \cite{Choudhury:2012yh,Choudhury:2012kw,Berg:2005ja}. The total phenomenological model is
made up of the following crucial components:
\begin{itemize}
 \item  Higher order correction terms in the gravity sector are introduced in the effective action as a perturbative correction to
 the Einstein-Hillbert counterpart coming 
from the computation of Conformal Field Theory disk amplitude at the two loop level \cite{Choudhury:2013yg,Choudhury:2013dia,Choudhury:2013eoa}.
\item The matter sector encounters the effect of ${\cal N}=1,{\cal D}=4$ supergravity motivated DBI Galileon interaction which is embaded in the D3 brane.
\item Additionally we have considered the effect of R-parity violating interactions \cite{Hambye:1999pw,Hambye:2000zs,Dreiner:1997uz,Martin:1997ns} in the matter sector which provide a convenient framework for quantifying quark
and lepton-flavor violating effects.
\end{itemize}

 The low energy UV protective effective action for
the proposed cosmophenomenological model is described by \cite{Choudhury:2012yh,Choudhury:2012kw}:
\begin{widetext}
\be\begin{array}{lll}\label{eq1}
    \displaystyle S=\int d^{4}x \sqrt{-g}
\left[K({\bf \Phi},X)-G({\bf \Phi},X)\Box{\bf \Phi}
+{\cal B}_{1}R+\left({\cal B}_{2}R^{\alpha\beta\gamma\delta}R_{\alpha\beta\gamma\delta}
-4{\cal B}_{3}R^{\alpha\beta}R_{\alpha\beta}
+{\cal B}_{4}R^{2}\right)+{\cal B}_{5}\right]
   \end{array}\ee
\end{widetext}
where the model dependent characteristic functions $K({\bf \Phi},X)$ and $G({\bf \Phi},X)$ are the implicit functions
 of galileon and its kinetic counterpart is $X=-\frac{1}{2}g^{\mu\nu}\partial_{\mu}{\bf \Phi}\partial_{\nu}{\bf \Phi}$.
Additionally, ${\cal B}_{i}\forall i$ are the self-coupling constants of graviton degrees of freedom appearing via dimensional reduction from higher dimensional string theory. Specifically 
${\cal B}_{5}$ be the effective four dimensional cosmological constant. In general, ${\cal B}_{2}\neq {\cal B}_{3} \neq {\cal B}_{4}$
which implies that the quadratic curvature terms originated from two loop correction to the CFT disk amplitudes are not topologically invariant in 4D effective theory. 
In Eq~(\ref{eq1}) for potential driven DBI Galileon model once we embed DBI theory in the Galileon background we can write, 
$K({\bf \Phi},X)=P({\bf \Phi},X)-V({\bf \Phi})$,
where the kinetic term of the effective action is given by:\begin{widetext}
   \be\begin{array}{llll}\label{plo}
   \displaystyle  P({\bf \Phi},X)=-\frac{{\cal G}_{1}}{f({\bf \Phi})}\left[\sqrt{1-2{\cal G}_{2}Xf({\bf \Phi})}-{\cal G}_{3}\right]
-{\cal G}_{4}G({\bf \Phi},X)-2{\cal G}_{5}X   
      \end{array}\ee
     \end{widetext}
with an effective {\it Klebanov Strassler } frame
function 
\be f({\bf \Phi})=(\sum^{2}_{q=0}f_{2q}{\bf \Phi}^{2q})^{-1}\ee 
which characterizes the throat geometry on the D3 brane. 
Here ${\cal G}_{i}\forall i$ and $f_{2q}\forall q$ are originated from dimensional reduction. 
 It is important to note that the fuctional $G(\phi,X)$ appearing in Eq~(\ref{eq1}) and Eq~(\ref{plo}) are exactly same in the context of DBI Galilon theory.
 For more details on this issue 
see Refs.~\cite{Choudhury:2012yh,Choudhury:2012kw}.
 In the canonical limit when the contributions from the DBI Galileon sector is switched off then we get, $K(\phi,X)=X-V(\phi)$, for ${\cal G}_{5}=-1/2$.
 In such a case the contributions from the higher derivative terms are highly suppressed by the {\bf UV} cut-off scale of the effective theory and finally 
we get back the usual results as obtained 
from Einstein gravity. But once the contribution of DBI Galileon is switched on, the complete analysis devaties from canonical behaviour and contribution from the 
higher derivative gravity sector plays crucial role to change the dynamical behaviour during inflation as well as reheating.

Moreover, the 
one-loop effective Coleman-Weinberg potential is given by \cite{Choudhury:2012yh,Choudhury:2012kw}:
\be\begin{array}{lll}\label{eq2m}\displaystyle V({\bf \Phi})=\sum^{2}_{m=-2,m\neq -1}\left[\beta_{2m}+\delta_{2m}\ln\left(\frac{{\bf \Phi}}{M}
\right)\right]\left(\frac{{\bf \Phi}}{M}\right)^{2m}\end{array}\ee
where $\beta_{2m}\forall m$ are the tree level constants and $\delta_{2m}\forall m$ are originated from one-loop correction.
In the present setup using Eq~(\ref{eq1}) the Modified Friedman Eqn can be expressed as \cite{Choudhury:2012yh,Choudhury:2012kw}:
\be\begin{array}{llll}\label{fred}\displaystyle H^{4}=\frac{\rho}{3M^{2}_{PL}\theta},\end{array}\ee
 where the energy density can be written in terms of DBI Galileon degrees of freedom as:
\be\begin{array}{llll}\label{rhodbi}\displaystyle\rho=2K_{X} X -K -2G_{\phi} X -2X(1-\theta)+V(\phi).\end{array}\ee 
 Here the subscipts represent the derivatives with respect to $X$ and $\phi$.
  Moreover, $\theta$ be a constant which is appearing through dimensional reduction from higher dimensional stringy setup as:
 \be\begin{array}{lll}\displaystyle \theta= \frac{{\cal B}_{1}}{M^{2}_{P}}+{\cal B}_{2}-4{\cal B}_{3}+{\cal B}_{4}+\frac{{\cal B}_{5}}{M^{4}_{P}},\end{array}\ee
In the present setup, $M_{P}=2.4\times 10^{18}~GeV$ be the reduced Planck mass.
  It is important to note 
 that the Firedmann Eqn obtained in the present context is completely
 different from the Friedman Eqn as appearing in the context of Einstein's General Relativity which will further modifies the leptogenesis
 framework in the present context.

To study this feature explicitly further we allow interaction of DBI Galileon scalar degrees of freedom with leptonic sector 
of the theory given by:
\be\begin{array}{lll}\label{eq2}
\displaystyle {\cal L}^{int}_{R\slashed}=\sum_{k}\left[{\cal Y}^{ijk}_{1}\nu_{i}l_{j}{\bf \Phi} +{\cal Y}^{ijk}_{2}\nu_{i}\bar{l}_{j}{\bf \Phi}+ h.c.\right]
   \end{array}\ee
where the generation indices are $i,j,k=1(e),2(\mu),3(\tau)$.
Here after summing over all the contributions of flavor indices the corresponding charged scalar field 
can be written as:
\be\begin{array}{llll}\label{eq3a}
    \displaystyle {\bf \Phi}=\frac{({\bf \Phi}_{e}\oplus{\bf \Phi}_{\mu}\oplus {\bf \Phi}_{\tau})}{\sqrt{3}}.
   \end{array}\ee
This induces the decay of charged DBI Galileon (${\bf \Phi}[{\bf \Phi}^{+},{\bf \Phi}^{-}]$) to the leptonic constituents through the phenomenological
 couplings $({\cal Y}^{ijk}_{1},{\cal Y}^{ijk}_{2})$. 
In this context these couplings violate a discrete symmetry called R-parity
defined as, $R_{p}\equiv (-1)^{3B+L+2S}$, where $B,L$ and $S$ are the baryon, lepton and spin angular momentum
respectively. Such R-parity violating interactions in the Lagrangian (\ref{eq2}) can be identified with the lepton number violating (LNV) MSSM flat direction ${\bf LLe}$ 
appearing in the superpotential as \cite{Dreiner:1997uz,Martin:1997ns}:
\be\begin{array}{lll}\label{eq3}
    \displaystyle {\cal W}^{MSSM}_{R\slashed}\supset \frac{1}{2}\epsilon_{ab} {\lambda}^{ijk}{\bf L}^{a}_{i}{\bf L}^{b}_{j}\bar{\bf e}_{k}+h.c.
   \end{array}\ee
where $a,b=1,2$ are weak isospin indices and flatness constraint requires $i<j$ for the lepton doublet {\bf L}. Here ${\lambda}^{ijk}$
be the soft SUSY breaking trilinear coupling which violate the R-parity and proton-hexality $P_{6}$, however,
they conserve baryon triality $B_{3}$ \cite{Dreiner:2010ye,Dreiner:2005rd,Banks:1991xj}. Additionally, due to the large suppression of the
baryon number violating interactions via $B_{3}$ triality it stabilizes the proton. 
Now using the constraints on mass $m_{\nu_{j}}<<m_{{\bf \Phi}}$ the corresponding decay widths for the feasible decay channels are:
\be\begin{array}{llll}\label{asq1} 
\displaystyle \Gamma({\bf \Phi}^{+}\rightarrow\nu_i l^{+}_j) = \frac{m_{{\bf \Phi}}}{16\pi^2}{\bf \tilde{F}_{ij}},\\ \\
\displaystyle \Gamma({\bf \Phi}^{-}\rightarrow\nu_i l^{-}_j) = \frac{m_{\bf \Phi}}{16\pi^2}{\bf \tilde{B}_{ij}}. \end{array}\ee
where the bilinear functions ${\bf \tilde{F}_{ij}}$ and ${\bf \tilde{B}_{ij}}$, can be expressed as:
\be\begin{array}{llll}\label{biln}\displaystyle m_{\bf \Phi}{\bf \tilde{F}_{ij}}=\frac{1}{3} \sum_{k}m_{\phi_k}{\bf F_{ijk}}, \\
\displaystyle m_{\bf \Phi}{\bf \tilde{B}_{ij}} = \frac{1}{3}\sum_{k} m_{\phi_k}{\bf B_{ijk}} .\end{array}\ee

\begin{figure}[t]
\centering
\includegraphics[width=12.5cm,height=8cm]{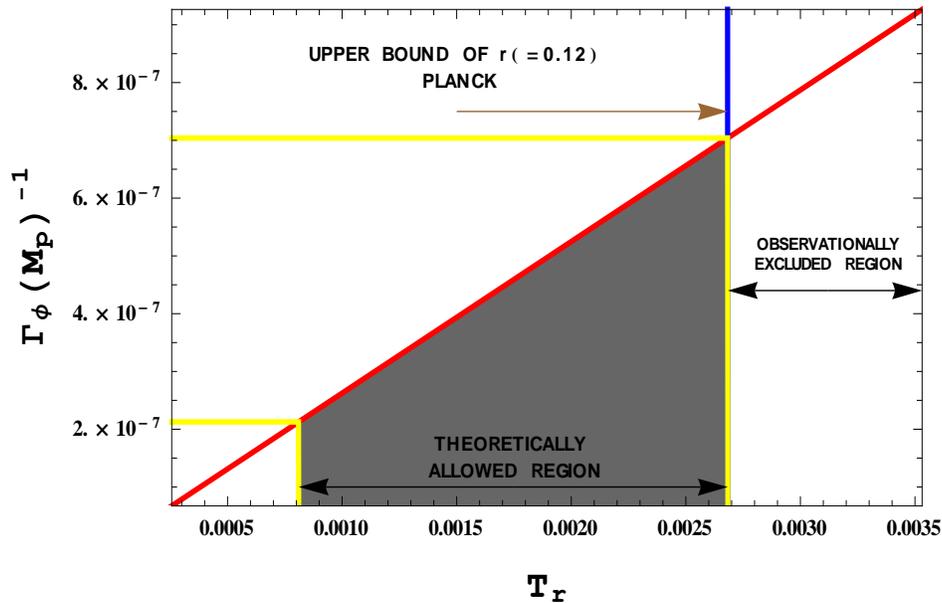}
\caption{Variation of total decay width $\Gamma_{\phi}$ 
 with respect to reheating temperature ($T_{r}$). The dark gray shaded area shows the theoretically allowed region which lies within the upper bound of
the tensor-to-scalar ratio ($r_{\star}\leq 0.12$) at the momentum pivot scale $k_{\star}=0.002~Mpc^{-1}$ represented by a blue vertical line in light of PLANCK data.
 We have also pointed the excluded parameter space for the tensor-to-scalar ratio within the range $0.12<r_{\star}\leq 0.36$ by imposing the constraint from PLANCK data.
For the numerical estimation in the present context we have used,
${\cal G}_{1}=1$, ${\cal G}_{2}=0.5$, ${\cal G}_{3}=2$, ${\cal G}_{4}=1$, ${\cal G}_{5}=-0.5$
and ${\cal B}_{1}=2M^{2}_{P}$,
     ${\cal B}_{2}=2$, ${\cal B}_{3}=1$, 
     ${\cal B}_{4}=3$, 
     ${\cal B}_{5}=2M^{4}_{P}$.
}
\label{f2}
\end{figure}

Here $m_{\bf \Phi}$ be the flavour independent inflaton mass and $m_{\phi_k}$ represents the k-th flavour dependent mass of the constituent ${\bf \Phi}_{k}$.
Additionally, the expression for trilinear functions ${\bf F_{ijk}}$ and ${\bf B_{ijk}}$ are explicitly mentioned in the appendix.
   To understand the thermal history of the universe from our model, it is convenient to express the decay width in terms of the Hubble parameter during the 
epoch of reheating as:
\be\begin{array}{lll}\label{eq5}
  \displaystyle  m_{\bf \Phi}\sum_{ij}\left( \displaystyle {\bf \tilde{F}_{ij}}+{\bf \tilde{B}_{ij}} \right) = 16\pi^2\Gamma_{\phi}(T_{r})=48\pi H_{rh}
   \end{array}\ee
where $\Gamma_{\phi}(T_{r})$ be the total decay width. In the present context the Hubble parameter during reheating is defined through the modified Friedmann equation as given by:
\be H_{rh}\approx\sqrt[4]{\frac{\rho_{rh}}{3M^{2}_{P}\theta}}.\ee
Here $\rho_{rh}$ be the energy density 
during reheating. Hence using eqn~(\ref{eq5}) the reheating
temperature can be expressed as:
\begin{widetext}
\be\begin{array}{llll}\label{eq6}
    \displaystyle T_{r}= \sum_{ij} \sqrt[4]{\frac{5M^{2}_{P}\theta~ m^{4}_{{\bf \Phi}}}{294912\pi^{6}N_{\star}}}({\bf \tilde{F}_{ij}}+{\bf \tilde{B}_{ij}})
=\sqrt[4]{\frac{5M^{2}_{P}\theta}{18432\pi^{2}N_{\star}}}\Gamma_{\phi}
   \end{array}\ee
   \end{widetext}
where $N_{\star}(=N_{B}+\frac{7}{8}N_{F})$ be the effective number of relativistic degrees of freedom. Usually $N_{\star}\approx 228.75$ for
all MSSM degrees of freedom. Recent observational data from PLANCK suggests an upper-bound on the reheating temperature \cite{Choudhury:2013jya,Choudhury:2013iaa,
Choudhury:2013woa,Ade:2013uln}:
\be\begin{array}{lll}\label{eq7}
    \displaystyle T_{r}\leq 6.654\times 10^{15}\sqrt[4]{\frac{r_{\star}}{0.12}}~GeV
   \end{array}\ee
 where $r_{\star}$ be the tensor-to-scalar ratio at the pivot scale of momentum $k_{\star}$. Consequently
the upper-bound of total decay width during reheating is given by:
\begin{widetext}
\be\begin{array}{lllll}\label{eq8}
    \displaystyle \Gamma_{\phi}=\sum_{ij} \frac{m_{{\bf \Phi}}}{16\pi^2}({\bf \tilde{F}_{ij}}+{\bf \tilde{B}_{ij}})
\leq 2.772\times 10^{-3}\sqrt[4]{\frac{3072\pi^{2}M^{2}_{P}N_{\star}r_{\star}}{\theta}}
   \end{array}\ee
 \end{widetext} 
where the stringent constraint on the slepton masses and soft SUSY breaking triliear coupling are $m_{\tilde{e}_k} = 300 ~ GeV$ and $\sum_{ijk}|\lambda_{ijk}|^2 = 1.7786$  
at the GUT scale, which are obtained by solving the one-loop renormalization group equation in ${\bf\overline{DR}}$ scheme \cite{Dreiner:2010ye}. In fig~(\ref{f2}) we
have shown the behaviour of the total decay width as a function of reheating temperature by imposing the observational constraints in light of PLANCK data. Additionally
we have also pointed the theoretically allowed region obtained from the model as well as the observationally excluded parameter space.

It is important to note that saturating the upper-bound on $r\sim 0.12$ would yield a large reheating temperature of the universe. 
In this case, the gravitino abundance is compatible with the latest obseravational/phenomenological bound on dark matter, provided the gravitino mass,
$m_{3/2}\sim {\cal O}(100)$~eV, see~\cite{Ibe:2010ym}. The light gravitino is a
very interesting candidate for dark matter among various other candidates, since the gravitino itself is a
unique and inevitable prediction of supergravity (SUGRA) theory. This prediction is very much
interesting, since we can test the gravitino dark matter hypothesis at LHC or through any other indirect probes. 
In fact, if we had late time entropy production after the decoupling
time of the gravitino, the mass of the gravitino dark matter may be raised up to a few
keV. Moreover, the gravitino dark matter with a mass in the range $m_{3/2}\sim {\cal O}(1-10)$~keV serves
as the warm dark matter candidate which has recently been invoked as possible solutions to the
seeming discrepancies between the observation and the simulated results of the galaxy
formation based on the cold dark matter (CDM) scenario \cite{deVega:2010yk,Markovic:2010te,VillaescusaNavarro:2010qy,Boyanovsky:2010sv}.
 See \cite{Ibe:2010ym} for the deatils of such scenario.
Additionally, the
gravitino mass of this order is also favored from several other phenomenological issues, the
interesting parameter space for the gaugino masses at the LHC, and the solution to the well known $\mu$-problem \cite{Yanagida:1997yf}.

By assuming such a phenomenological prescription perfectly holds good in our prescribed string theoretic setup
let us start with a situation where the inflaton field starts
oscillating when the inflationary epoch ends at a cosmic time
$t=t_{osc}\simeq t_{f}$ and the reheating phenomenology is described
by the {\it Boltzmann equation} \cite{Choudhury:2011rz}:
\be\label{rfg} \dot{\rho_{r}}+4H\rho_{r}=\Gamma_{\phi}\rho_{\phi},\ee
where 
$H\approx\sqrt[4]{\frac{\rho_{r}+\rho_{\phi}}{3M^{2}_{P}\theta}}$.
Here $\rho_{r}$ and $\rho_{\phi}$ represent the energy density of radiation and inflaton respectively. Assuming $\Gamma_{\phi}\gg H$ from we get \be\label{gh1}
\rho_{\phi}=\frac{\rho_{\phi_0}}{x^4}\exp\left[-\Gamma_{\phi}(t-t_{osc})\right]\ee
where $\rho_{\phi_0}=\beta_{0}$ (the energy scale of DBI Galileon inflation as appearing in Eq~(\ref{eq2m})) and additionally we introduce a new parameter ``x" defined as:
\be\begin{array}{lll}\label{xcd}
    \displaystyle x:=\frac{a}{a_{osc}}=\left[1+H_{osc}(t-t_{osc})\right]
   \end{array}\ee
with  $H_{osc}=\left(\frac{\rho_{osc}}{3M_p^2\theta}\right)^{1/4}$.
For $t\leq \Gamma^{-1}_{\phi}$ the exact solution of the eqn(\ref{rfg}) can be written as \bea\label{sd1}
        \rho_{r} &=& \frac{1}{x^4}\left[\rho_{osc}-\rho_{\phi_{0}}\exp\left(-\frac{(x-1)\Gamma_{\phi}}{H_{osc}}\right)\right].
\eea
 
Finally we are interested in to compute the thermal dark matter 
gravitino relic abundance produced by the scattering of the inflaton decay
 products. To serve this purpose we start with the master equation of gravitino
 phenomenology as obtained from {\it Boltzmann equation}
 is given by \cite{Choudhury:2011rz}:
  \be\label{bn1}
\left(\frac{d}{dt}+3H\right)n_{3/2}=\langle\Sigma_{total}|v|\rangle n^{2}
-\frac{m_{3/2}n_{3/2}}{\langle E_{3/2}\rangle\tau_{3/2}},\ee
 where $n=\frac{\zeta(3)T^{3}}{\pi^{2}}$ is the number density of scatterers(bosons in thermal bath) with
$\zeta(3)$=1.20206.... Here $\Sigma_{total}$ is the total scattering cross section for thermal gravitino production, $v$ is the relative velocity of the incoming particles with $\langle v\rangle=1$ where $\langle...\rangle$ represents the thermal average. The factor $\frac{m_{\frac{3}{2}}}{\langle E_{\frac{3}{2}}\rangle}$ represents the averaged Lorentz factor which comes from the decay of gravitinos can be neglected due to weak interaction.
 For the gauge group $SU(3)_{C}\otimes SU(2)_{L}\otimes U(1)_{Y}$ the thermal gravitino production rate is given by, \be\begin{array}{ll}\label{mn1}
\displaystyle\langle\Sigma_{total}|v|\rangle=\frac{\tilde{\alpha}}{M^{2}_{P}}
\\~~~~~~~~~~~~~\displaystyle=\frac{3\pi}{16\zeta(3)M^{2}_{P}}\sum^{3}_{i=1}\left[1+\frac{M^{2}_{i}}{3m^{2}_{3/2}}\right]
C_{i}g^{2}_{i}\ln\left(\frac{K_{i}}{g_{i}}\right),\end{array}\ee where $i=1,2,3$ stands for the three
 gauge groups $U(1)_{Y}$,$SU(2)_{L}$ and $SU(3)_{C}$ respectively. Here $M_{i}$ represent gaugino mass parameters and $g_{i}(T)$ represents
 gaugino coupling constant at finite temperature (from MSSM RGE)\cite{Choudhury:2011rz,Choudhury:2012ib}:\be\label{hh1}
g_{i}(T)\simeq\frac{1}{\sqrt{\frac{1}{g^{2}_{i}(M_{Z})}-\frac{b_{i}}{8\pi^{2}}\ln\left(\frac{T}{M_{Z}}\right)}}\ee
with $b_{1}=11,b_{2}=1,b_{3}=-3$. Here $C_{i}$ and $K_{i}$ represents the constant
 associated with the gauge groups $U(1)_{Y}$,$SU(2)_{L}$ and $SU(3)_{C}$ with $C_{1}=11,C_{2}=27,C_{3}=72$ and $K_{1}=1.266,K_{2}=1.312,K_{3}=1.271$ \cite{Choudhury:2011rz,Choudhury:2012ib}.
 
Further re-expressing Eq~(\ref{bn1}) in terms of the parameter ``x'' and imposing 
 the boundary condition $\dot{T}=0$ at maximum energy density $\rho_{r}(x=x_{max})$ the thermal gravitino dark matter relic abundance is given by
\begin{widetext}
\be
\begin{array}{ll}
\label{d1}\displaystyle \Omega_{3/2}(x)=\frac{n_{3/2}(x)}{s(x)}
=\frac{45}{2\pi^2x^3 N_{\star}T^3(x)}\left[\frac{C_{3}}{4}\sqrt{\frac{C_{2}+C_{1}(x-1)}{x^4}}\left\{
C_{1}(2-5x)-2C_{2}\right.\right.\\ \left.\left. \displaystyle ~~~~~~~~~~~~~~~~~~~~~~~~~~~~~~~~~~~~~~~~~~~~~~~~~~~~~~~~
+\frac{3C^{2}_{1}x^2}{\sqrt{(C_{1}-C_{2})(C_{2}+C_{1}(x-1))}}
tan^{-1}\left(\sqrt{\frac{C_{2}+C_{1}(x-1)}{C_{1}-C_{2}}}\right)\right\}\right.\\ \left. 
 \displaystyle ~~~~~~~~~~~~~~~~~~~~~~~~~~~~~~~~~+\frac{1}{3x^2_{max}}\left\{C_{3}(C_{2}+C_{1}(x_{max}-1))\right\}^{\frac{3}{2}}-\frac{C_{3}}{4}\sqrt{\frac{C_{2}+C_{1}(x_{max}-1)}{x^4_{max}}}\left\{
C_{1}(2-5x_{max})-2C_{2}\right.\right.\\ \left.\left.\displaystyle~~~~~~~~~~~~~~~~~~~~~~~~~~~~~~~~~~~~~~~~~~~~~+\frac{3C^{2}_{1}x^2_{max}}{\sqrt{(C_{1}-C_{2})(C_{2}+C_{1}(x_{max}-1))}}
tan^{-1}\left(\sqrt{\frac{C_{2}+C_{1}(x_{max}-1)}{C_{1}-C_{2}}}\right)\right\}\right],
\end{array}
\ee
\end{widetext}

\begin{figure*}[htb]
\centering
\includegraphics[width=12.5cm,height=8cm]{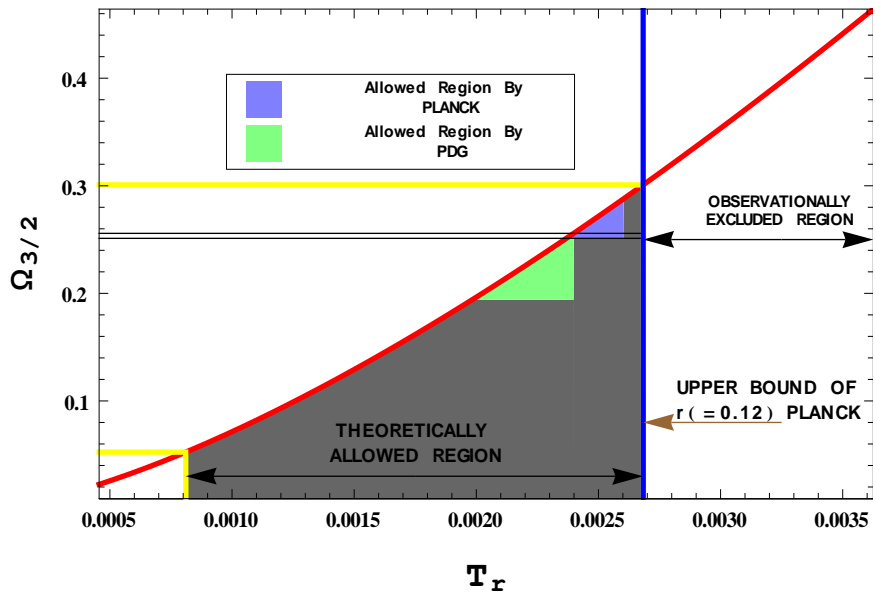}
\caption{Variation of gravitino dark matter relic density parameter ($\Omega_{3/2}$) with respect to reheating temperature ($T_{r}$).
The dark gray shaded region shows the theoretically allowed region which lies within the upper bound of
the tensor-to-scalar ratio ($r_{\star}\leq 0.12$) at the momentum pivot scale $k_{\star}=0.002~Mpc^{-1}$ represented by a blue vertical line in light of PLANCK data. 
We have explicitly shown the observationally allowed region of $\Omega_{3/2}$ by imposing the constraints from the
PLANCK data. Further we have also pointed the constraint parameter space obtained from the PDG catalog. Most importantly, 
the overlapping region within the range $0.245<\Omega_{3/2}<0.250$ shown by the black strip satisfies both the constraints obtained from PLANCK and PDG data. 
For the numerical estimation in the present context we have used,
${\cal G}_{1}=1$, ${\cal G}_{2}=0.5$, ${\cal G}_{3}=2$, ${\cal G}_{4}=1$, ${\cal G}_{5}=-0.5$
and ${\cal B}_{1}=2M^{2}_{P}$,
     ${\cal B}_{2}=2$, ${\cal B}_{3}=1$, 
     ${\cal B}_{4}=3$, 
     ${\cal B}_{5}=2M^{4}_{P}$.
}
\label{f3}
\end{figure*}
 where the entropy density is given by $s(x)=\frac{2\pi^2}{45}N_{\star}T^3(x)$. Here the 
temperature can be expressed in terms of the tensor-to-scalar ratio and the parameter ``x'' as:
\begin{widetext}

\be\begin{array}{lll}\label{tempra}
    \displaystyle T(x)=\frac{1}{\sqrt{\pi}}\sqrt[4]{\left[\frac{30}{N_{\star}x^4}\left(\rho_{osc}-\rho_{\phi_{0}}\left\{1-\frac{0.48(x-1)}{H_{osc}}
\sqrt[4]{\left(\frac{N_{\star}M^2_{P}}{\theta}\right)}\right\}\right)\right]}
   \end{array}\ee
\end{widetext}

and we also introduce new sets of parameters defined as:
\be\begin{array}{lll}\label{zse1}
    \displaystyle C_{1}=\frac{\Gamma_{\phi}\rho_{\phi_{0}}}{H_{osc}},~~~
\displaystyle C_{2}=\rho_{osc}-\rho_{\phi_{0}},\\
\displaystyle C_{3}=\frac{30\sqrt{10}\tilde{\alpha}\zeta^{2}(3)}{H_{osc}\pi^{7}M^2_{P} N^{\frac{3}{2}}_{\star}},~~~
\displaystyle x_{max}=\frac{4}{3}-\frac{4C_{2}}{9C_{1} H^{2}_{osc}}.
   \end{array}\ee

In this paper we introduce the leptogenesis scenario in presence of 
 DBI Galileon which has the following remarkable phenomenological features:

\begin{itemize}
 \item In Fig~(\ref{f2}), the
 theoretically allowed region shows that the reheating tempreature for DBI Galileon
 is high enough and lies around the GUT scale ($10^{16}$~GeV). This is the first
 observation we have made from our analysis in the context of DBI Galileon, which
 is remarkably diffrent from the GR prescribed setup as using GR we can probe upto $10^{10}$~GeV.
 Such high values of the reheating temperature implies that
 the obtained value of the tensor-to-scalar ratio 
 from the DBI Galileon inflationary set up lies within a wide range: $2.4\times 10^{-3}<r_{\star}<0.12$, at the piviot scale of momentum $k_{\star}\sim 0.002~{\rm Mpc}^{-1}$,
 which confronts well the Planck data.
 If the signatures of the primordial gravity waves will be detected at present or in near future then the consistency
 between the high rehating temperature and garvity waves can be directly verified from our prescribed model using Eq~(\ref{eq7}).

\item 
In fig(\ref{f3}) we have explicitly shown the behaviour of gravitino relic abundance with respect to reheating temperature 
in light of PLANCK and PDG data. The overlapping region
 within the range $0.245<\Omega_{3/2}<0.250$ satisfies both the dark matter constraints obtained from PLANCK and PDG data as given by \cite{Ade:2013zuv,Ber:2012}:
\be\begin{array}{lll}\label{omeg}
    \displaystyle \Omega^{PLANCK}_{DM}=0.26\pm 0.01\\
\displaystyle ~~~\Omega^{PDG}_{DM}= 0.22\pm 0.03.
   \end{array}\ee

\end{itemize}
 
In the present article we have studied cosmological consequences of reheating and dark matter phenomenology in the context of DBI Galileon 
on the background of low energy effective supergravity framework. We have engaged ourselves in investigating for the effect of perturbative reheating
by imposing the constraints from primordial gravitational waves in light of the PLANCK data. Further 
we have established a cosmological connection between thermal gravitino dark matter relic abundance, reheating temperature
and tensor-to scalar ratio in the present context. To this end we have 
explored the model dependent features of thermal relic gravitino abundance 
by imposing the dark matter constraint from PLANCK+PDG data, which is 
also consistent with the additional constraint associated with the upper bound of tensor-to-scalar ratio $r_{0.002}\leq 0.12$ obtained from PLANCK data.
.


\section*{Acknowledgments}

SC thanks Council of Scientific and
Industrial Research, India for financial support through Senior
Research Fellowship (Grant No. 09/093(0132)/2010). SC also thanks 
Centre for Theoretical Physics, Jamia Millia Islamia for extending hospitality.
AD thanks Council of Scientific and
Industrial Research, India for financial support through Senior
Research Fellowship (Grant No. 09/466(0125)/2010).
\newline
\section*{Appendix}
In Eq~(\ref{asq1},\ref{eq5},\ref{eq6},\ref{eq8}) the trilienar functions are given by:
\bea\label{cv1}
   {\bf F_{ijk}} &=& |\lambda_{ijk}|^2\left[2 + \frac{9g^4}{4c^4_w}\left(\frac{1}{4}+2s^4_w-s^2_w\right)I^2_1(m^2_{\phi_k},0,0)\right] \nonumber \\
   {\bf B_{ijk}} &=& |\lambda_{ijk}|^2\left[2 + \frac{9g^4}{4c^4_w}\left(\frac{1}{4}+2s^4_w+s^2_w\right)I^2_1(m^2_{\phi_k},0,0)\right] \nonumber \\
   \eea
where the integral $I_1(m^2_k,m^2_i,m^2_j)$ is defined as
\begin{widetext}
\bea\label{cvcx1}
I_1(m^2_k,m^2_i,m^2_j) &=& -i \int_0^1\int_0^1 dx dy \Big{[}\gamma_{E} + \frac{1}{2} 
+ \frac{N(x,y)}{2Q^2(x,y)}+\ln\left[\frac{Q^2(x,y)}{4\pi \mu^2}\right]\Big{]} \eea
with
\bea\label{cvcx2}
N(x,y) &=& x(1-x)m^2_j + y(1-y)m^2_i + \frac{1}{2}\left[(1-x-y)(m^2_k-m^2_i-m^2_j)\right] \nonumber \\
Q^2(x,y) &=& x^2m^2_j + y^2m^2_i -xy (m^2_k-m^2_i-m^2_j) + (1-x-y)m^2_z. \nonumber \\
\eea
\end{widetext}
In Eq~(\ref{cvcx1}) $\gamma_{E}= 0.5772$ is the 
{\it Euler-Mascheroni constant} originating in the expansion of the gamma function. Here $c_w=cos\theta_w, s_w=sin\theta_w$ (where $\theta_w$= Weinberg angle) 
and $m_z$ be the mass of the Z boson.




\begin{references}

\bibitem{Allahverdi:2007zz}
  R.~Allahverdi and A.~Mazumdar,
  Phys.\ Rev.\ D {\bf 76} (2007) 103526
  [hep-ph/0603244].

\bibitem{Allahverdi:2010xz}
  R.~Allahverdi, R.~Brandenberger, F.~-Y.~Cyr-Racine and A.~Mazumdar,
  Ann.\ Rev.\ Nucl.\ Part.\ Sci.\  {\bf 60} (2010) 27
  [arXiv:1001.2600 [hep-th]].

\bibitem{Allahverdi:2006we}
  R.~Allahverdi, K.~Enqvist, J.~Garcia-Bellido, A.~Jokinen and A.~Mazumdar,
  JCAP {\bf 0706} (2007) 019
  [hep-ph/0610134].


\bibitem{Cardenas:2007xh}
  V.~H.~Cardenas,
  Phys.\ Rev.\ D {\bf 75} (2007) 083512
  [astro-ph/0701624].

\bibitem{Boyanovsky:1995ema}
  D.~Boyanovsky, M.~D'Attanasio, H.~J.~de Vega, R.~Holman and D.~-S.~Lee,
  Phys.\ Rev.\ D {\bf 52} (1995) 6805
  [hep-ph/9507414].

\bibitem{Kofman:1997yn}
  L.~Kofman, A.~D.~Linde and A.~A.~Starobinsky,
  Phys.\ Rev.\ D {\bf 56} (1997) 3258
  [hep-ph/9704452].

\bibitem{McDonald:1999hd}
  J.~McDonald,
  Phys.\ Rev.\ D {\bf 61} (2000) 083513
  [hep-ph/9909467].

\bibitem{Fong:2013wr}
  C.~S.~Fong, E.~Nardi and A.~Riotto,
  Adv.\ High Energy Phys.\  {\bf 2012} (2012) 158303
  [arXiv:1301.3062 [hep-ph]].

\bibitem{Okada:2005kv}
  N.~Okada and O.~Seto,
  Phys.\ Rev.\ D {\bf 73} (2006) 063505
  [hep-ph/0507279].

\bibitem{Pradler:2006qh}
  J.~Pradler and F.~D.~Steffen,
  Phys.\ Rev.\ D {\bf 75} (2007) 023509
  [hep-ph/0608344].

\bibitem{Rangarajan:2006xg}
  R.~Rangarajan and N.~Sahu,
  Mod.\ Phys.\ Lett.\ A {\bf 23} (2008) 427
  [hep-ph/0606228].

\bibitem{Mazumdar:2010sa}
  A.~Mazumdar and J.~Rocher,
  Phys.\ Rept.\  {\bf 497} (2011) 85
  [arXiv:1001.0993 [hep-ph]].


\bibitem{Enqvist:2003gh}
  K.~Enqvist and A.~Mazumdar,
  Phys.\ Rept.\  {\bf 380} (2003) 99
  [hep-ph/0209244].

\bibitem{Jungman:1995df}
  G.~Jungman, M.~Kamionkowski and K.~Griest,
  Phys.\ Rept.\  {\bf 267} (1996) 195
  [hep-ph/9506380].

\bibitem{Pradler:2006hh}
  J.~Pradler and F.~D.~Steffen,
  Phys.\ Lett.\ B {\bf 648} (2007) 224
  [hep-ph/0612291].


\bibitem{Pospelov:2010hj}
  M.~Pospelov and J.~Pradler,
  Ann.\ Rev.\ Nucl.\ Part.\ Sci.\  {\bf 60} (2010) 539
  [arXiv:1011.1054 [hep-ph]].

\bibitem{Burles:2000ju}
  S.~Burles, K.~M.~Nollett and M.~S.~Turner,
  Phys.\ Rev.\ D {\bf 63} (2001) 063512
  [astro-ph/0008495].

\bibitem{Kohri:2005wn}
  K.~Kohri, T.~Moroi and A.~Yotsuyanagi,
  Phys.\ Rev.\ D {\bf 73} (2006) 123511
  [hep-ph/0507245].

\bibitem{Kallosh:1999jj}
  R.~Kallosh, L.~Kofman, A.~D.~Linde and A.~Van Proeyen,
  Phys.\ Rev.\ D {\bf 61} (2000) 103503
  [hep-th/9907124].

\bibitem{Maroto:1999ch}
  A.~L.~Maroto and A.~Mazumdar,
  Phys.\ Rev.\ Lett.\  {\bf 84} (2000) 1655
  [hep-ph/9904206].

\bibitem{Bolz:2000fu}
  M.~Bolz, A.~Brandenburg and W.~Buchmuller,
  Nucl.\ Phys.\ B {\bf 606} (2001) 518
   [Erratum-ibid.\ B {\bf 790} (2008) 336]
  [hep-ph/0012052].


\bibitem{Nilles:1983ge}
  H.~P.~Nilles,
  Phys.\ Rept.\  {\bf 110} (1984) 1.


\bibitem{Choudhury:2011sq}
  S.~Choudhury and S.~Pal,
  Phys.\ Rev.\ D {\bf 85} (2012) 043529
  [arXiv:1102.4206 [hep-th]].


\bibitem{Choudhury:2011rz}
  S.~Choudhury and S.~Pal,
  Nucl.\ Phys.\ B {\bf 857} (2012) 85
  [arXiv:1108.5676 [hep-ph]].

\bibitem{Choudhury:2012ib}
  S.~Choudhury and S.~Pal,
  J.\ Phys.\ Conf.\ Ser.\  {\bf 405} (2012) 012009
  [arXiv:1209.5883 [hep-th]].

\bibitem{Choudhury:2013jya}
  S.~Choudhury, A.~Mazumdar and S.~Pal,
  JCAP {\bf 07} (2013) 041
  [arXiv:1305.6398 [hep-ph]].

\bibitem{Choudhury:2013zna}
  S.~Choudhury, T.~Chakraborty and S.~Pal,
  arXiv:1305.0981 [hep-th].

\bibitem{Choudhury:2014sxa}
  S.~Choudhury, A.~Mazumdar and E.~Pukartas,
  arXiv:1402.1227 [hep-th].

\bibitem{Choudhury:2014uxa}
  S.~Choudhury,
  arXiv:1402.1251 [hep-th].


\bibitem{Mazumdar:2011ih}
  A.~Mazumdar, S.~Nadathur and P.~Stephens,
  Phys.\ Rev.\ D {\bf 85} (2012) 045001
  [arXiv:1105.0430 [hep-th]].


\bibitem{Choudhury:2012yh}
  S.~Choudhury and S.~Pal,
  Nucl.\ Phys.\ B {\bf 874} (2013) 85
  [arXiv:1208.4433 [hep-th]].

\bibitem{Choudhury:2012kw}
  S.~Choudhury and S.~Pal,
  arXiv:1210.4478 [hep-th].

\bibitem{Berg:2005ja}
  M.~Berg, M.~Haack and B.~Kors,
  JHEP {\bf 0511} (2005) 030
  [hep-th/0508043].


\bibitem{Choudhury:2013yg}
  S.~Choudhury and S.~Sengupta,
  JHEP {\bf 1302} (2013) 136
  [arXiv:1301.0918 [hep-th]].

\bibitem{Choudhury:2013dia}
  S.~Choudhury and S.~SenGupta,
  arXiv:1306.0492 [hep-th].

\bibitem{Choudhury:2013eoa}
  S.~Choudhury, S.~Sadhukhan and S.~SenGupta,
  arXiv:1308.1477 [hep-ph].

\bibitem{Hambye:1999pw}
  T.~Hambye, E.~Ma and U.~Sarkar,
  Phys.\ Rev.\ D {\bf 62} (2000) 015010
  [hep-ph/9911422].

\bibitem{Hambye:2000zs}
  T.~Hambye, E.~Ma and U.~Sarkar,
  Nucl.\ Phys.\ B {\bf 590} (2000) 429
  [hep-ph/0006173].

\bibitem{Dreiner:1997uz}
  H.~K.~Dreiner,
  In *Kane, G.L. (ed.): Perspectives on supersymmetry II* 565-583
  [hep-ph/9707435].

\bibitem{Martin:1997ns}
  S.~P.~Martin,
  In *Kane, G.L. (ed.): Perspectives on supersymmetry II* 1-153
  [hep-ph/9709356].

\bibitem{Dreiner:2010ye}
  H.~K.~Dreiner, M.~Hanussek and S.~Grab,
  Phys.\ Rev.\ D {\bf 82} (2010) 055027
  [arXiv:1005.3309 [hep-ph]].

\bibitem{Dreiner:2005rd}
  H.~K.~Dreiner, C.~Luhn and M.~Thormeier,
  Phys.\ Rev.\ D {\bf 73} (2006) 075007
  [hep-ph/0512163].


\bibitem{Banks:1991xj}
  T.~Banks and M.~Dine,
  Phys.\ Rev.\ D {\bf 45} (1992) 1424
  [hep-th/9109045].

\bibitem{Choudhury:2013iaa}
  S.~Choudhury and A.~Mazumdar,
  arXiv:1306.4496 [hep-ph].

\bibitem{Choudhury:2013woa}
  S.~Choudhury and A.~Mazumdar,
  arXiv:1307.5119 [astro-ph.CO].

\bibitem{Ade:2013uln}
  P.~A.~R.~Ade {\it et al.}  [Planck Collaboration],
  arXiv:1303.5082 [astro-ph.CO].


\bibitem{Ibe:2010ym}
  M.~Ibe, R.~Sato, T.~T.~Yanagida and K.~Yonekura,
  JHEP {\bf 1104} (2011) 077
  [arXiv:1012.5466 [hep-ph]].


\bibitem{deVega:2010yk}
  H.~J.~de Vega, P.~Salucci and N.~G.~Sanchez,
  New Astron.\  {\bf 17} (2012) 653
  [arXiv:1004.1908 [astro-ph.CO]].

\bibitem{Markovic:2010te}
  K.~Markovic, S.~Bridle, A.~Slosar and J.~Weller,
  JCAP {\bf 1101} (2011) 022
  [arXiv:1009.0218 [astro-ph.CO]].

\bibitem{VillaescusaNavarro:2010qy}
  F.~Villaescusa-Navarro and N.~Dalal,
  JCAP {\bf 1103} (2011) 024
  [arXiv:1010.3008 [astro-ph.CO]].

\bibitem{Boyanovsky:2010sv}
  D.~Boyanovsky,
  Phys.\ Rev.\ D {\bf 83} (2011) 103504
  [arXiv:1011.2217 [astro-ph.CO]].

\bibitem{Yanagida:1997yf}
  T.~Yanagida,
  Phys.\ Lett.\ B {\bf 400} (1997) 109
  [hep-ph/9701394].

\bibitem{Ade:2013zuv}
  P.~A.~R.~Ade {\it et al.}  [Planck Collaboration],
  arXiv:1303.5076 [astro-ph.CO].

\bibitem{Ber:2012}
 J. Beringer {\it et al.} (Particle Data Group), Phys.\ Rev.\ D {\bf 86}, 010001 (2012).














\end{references}
\end{document}